%% file: main.tex
\documentclass{sig-alternate}
\input{macros}

\begin{document}

\permission{}
\conferenceinfo{}{}
\copyrightetc{}
\crdata{978-1-4503-3433-4/15/04\$15.00\\
http://dx.doi.org/10.1145/2728606.2728634}

\title{Towards Personalized Prostate Cancer Therapy Using Delta-Reachability Analysis}%\titlenote{This work has been partially supported by award N00014-13-1-0090 of the US Office of Naval Research and award CNS0926181 of the National Science foundation (NSF).}}

\numberofauthors{6} 
\author{
% 1st. author
\alignauthor
Bing Liu\\
       \affaddr{School of Medicine}\\
       \affaddr{University of Pittsburgh}\\
       %\affaddr{Pittsburgh, PA, USA}\\
       \email{liubing@pitt.edu}
% 2nd. author
\alignauthor
Soonho Kong\\
       \affaddr{Computer Science Dept.}\\
       \affaddr{Carnegie Mellon University}\\
      % \affaddr{Pittsburgh, PA, USA}\\
       \email{soonhok@cs.cmu.edu}
% 3rd. author
\alignauthor 
Sicun Gao\\
       \affaddr{CSAIL}\\
       \affaddr{MIT}\\
       %\affaddr{Cambridge, MA, USA}\\
       \email{sicung@csail.mit.edu}
\and  % use '\and' if you need 'another row' of author names
% 4th. author
\alignauthor 
Paolo Zuliani\\
       \affaddr{School of Computer Science}\\
       \affaddr{Newcastle University}\\
      % \affaddr{Newcastle, UK}\\
       \email{paolo.zuliani@ncl.ac.uk}
% 5th. author
\alignauthor 
Edmund M. Clarke\\
       \affaddr{Computer Science Dept.}\\
       \affaddr{Carnegie Mellon University}\\
       %\affaddr{Pittsburgh, PA, USA}\\
       \email{emc@cs.cmu.edu}
}

\date{20 February, 2015}

\maketitle

\begin{abstract}
Recent clinical studies suggest that the efficacy of hormone therapy for prostate cancer depends on the characteristics of individual patients. In this paper, we develop a computational framework for identifying patient-specific androgen ablation therapy schedules for postponing the potential cancer relapse. We model the population dynamics of heterogeneous prostate cancer cells in response to androgen suppression as a nonlinear hybrid automaton. We estimate personalized kinetic parameters to characterize patients and employ $\delta$-reachability analysis to predict patient-specific therapeutic strategies. The results show that our methods are promising and may lead to a prognostic tool for prostate cancer therapy.
\end{abstract}

\category{D.2.4}{Software Engineering}{Software/Program Verification}[Model checking]
\category{J.3}{Life and Medical Sciences}{Biology and genetics}

\terms{Theory, Verification}

\keywords{hybrid systems, delta-reachability, systems biology, prostate cancer, personalized therapy}

\input{intro}
\input{model}
\input{method}

\input{results}

\input{conclusion}

%ACKNOWLEDGMENTS are optional
\section{Acknowledgments}
This work has been partially supported by award N00014-13-1-0090 of the US Office of Naval Research and award CNS0926181 of the National Science foundation (NSF).

\bibliographystyle{abbrv}
\bibliography{sigproc}

\end{document}

%% file: macros.tex
\usepackage{stmaryrd,amsmath,amssymb,newlfont,graphicx,verbatim}
\usepackage[ruled,lined,boxed,commentsnumbered,linesnumbered]{algorithm2e}
\usepackage{subfigure,url}
\usepackage{fancyvrb}

\newcounter{thedef} 

\newcounter{theass} 

\newcounter{theeg} 

\newcommand{\citep}{\cite}

\newcommand{\hide}[1]{}

\newcommand{\ie}{{\em i.e.}}

%% file: intro.tex
\section{Introduction}
Prostate cancer is the second leading cause of cancer-related deaths among men in United States. %\citep{cancerstat}. 
Hormone therapy in the form of androgen deprivation has been a cornerstone of the management of advanced prostate cancer for several decades. However, controversy remains regarding its optimal application \citep{nru}. Continuous androgen suppression (CAS) therapy has many side effects including anemia, osteoporosis, impotence, etc. Further, most patients experience a relapse after a median duration of 18-24 months of CAS treatment, due to the proliferation of castration resistant cancer cells (CRCs).

In order to reduce side effects of CAS and to delay the time to relapse, intermittent androgen suppression (IAS) was proposed to limit the duration of androgen-poor conditions and avoid emergence of CRCs \citep{bruchovsky95}. In particular, IAS therapy switches between on-treatment and off-treatment modes by monitoring the serum level of a tumor marker called prostate-specific antigen (PSA):

-- When the PSA level decreases and reaches a lower threshold value $r_0$, androgen suppression is suspended.

-- When the PSA level increases and reaches a upper threshold value $r_1$, androgen suppression is resumed by the administration of medical agents.

Recent clinical phase II and III trials confirmed that IAS has significant advantages in terms of quality of life and cost \cite{bruchovsky06,bruchovsky07}. However, with respect to time to relapse and cancer-specific survival, the clinical trials suggested that to what extent IAS is superior to CAS depends on the individual patient and the on- and off-treatment scheme \citep{bruchovsky06,bruchovsky07}. Thus, a crucial unsolved problem is how to design a personalized treatment scheme for each individual to achieve maximum therapeutic efficacy.

To answer this question, mathematical models have been developed to study the dynamics of prostate cancer under androgen suppression \citep{jackson04a,jackson04b,ideta08,hirata10,pnas11,portz12}. Recently, attempts have been made to computationally classify patients and obtain the optimal treatment scheme \citep{chaos10,suzuki10}. However, these results relied on simplifying nonlinear hybrid dynamical systems to more manageable versions such as piecewise linear models \citep{chaos10} and piecewise affine systems \citep{suzuki10}, which compromises the validity of the models. In this paper, we construct a nonlinear hybrid model to describe the prostate cancer progression dynamics under IAS thereapy. Our model extends the models previously proposed in \citep{jackson04a,jackson04b,ideta08}. We use $\delta$-reachability analysis to obtain the following results:

-- First, we show that our model is in good agreement with the published clinical data in literature \cite{ bruchovsky06,bruchovsky07}. It can depict the dynamical changes of proliferation rates induced by perturbing androgen levels that are difficult for previous models (e.g. \cite{ideta08}) to capture. It also addresses the variability in individual patients and is able to accurately reproduce the datasets of different patients.  

-- Second, we obtain interesting insights on CRC proliferation dynamics through analysis of the nonlinear model. Our results support the hypothesis that the physiological level of androgen reduce CRCs \cite{ideta08}, while rule out other hypotheses, for instance, CRCs proliferate at a constant rate \cite{portz12}. 

-- Third, we propose a computational framework for identifying patient-specific IAS schedules for postponing the potential cancer relapse. Specifically, we obtain personalized model parameters by fitting to the clinical data in order to characterize individual patients. We then use $\delta$-decision produces and bounded model checking to predict therapeutic strategies. 

Through this case study, we aim to highlight the opportunity for solving realistic biomedical problems using formal methods. In particular, methods based on $\delta$-reachability analysis suggest a very promising direction to proceed. \\

%% PLEASE CHECK

%\noindent \textit{Related Work.} Hybrid automata have been used to model cancer progression and drug effects \cite{bud14}. Of particular interest in our context are parameter synthesis for nonlinear hybrid automata. Tools such as SpaceEx \cite{spaceex} and Flow\* \cite{flowstar} can answer reachablitiy queries by computing an overappoximation of the reachable sets. However, it is difficult for them to perform parameter synthesis, as they cannot trace back to the parameter values that lead to the reachable region of interest. Rovergene tool \cite{rovergene} deals with parameter synthesis for piecewise affine linear models. Applying it in this case study requires linearization of the underlying nonlinear dynamics \cite{grosu11}. Alternatively, sampling/simulation based tools (e.g. Breach \cite{breach}) can also be used to perform parameter synthesis. Comparing to such tools, our approach explores the parameter space in a more efficient way, and can offers formal guarantees as well.

\noindent \textit{Related Work.} 
We perform parameter synthesis, which requires the computation of
concrete trajectories and parameter values. This can not be done by
simply computing an over-approximation of the forward reachable set.
Consequently, reachable set computation tools such as SpaceEx \cite{spaceex} and
Flow* \cite{flowstar} can not be directly used. There exists various approaches for
performing parameter synthesis through extra refinement on the
reachable sets~\cite{DreossiD14,BogomolovMPW14,FrehseJK08},
but are restricted to dynamics that are much simpler than the models
we encounter here. On the other hand, other SMT-based methods for
hybrid systems~\cite{CimattiGMT13,CimattiMT12},
which can perform parameter synthesis in a similar manner, mostly
focus on efficient handling of complex discrete transitions but are
restricted to models with simpler continuous dynamics.

The rest of the paper is organized as follows. We describe our model in Section 2 and present preliminaries on $\delta$-reachability analysis in Section 3. In Section 4, we present the biological insights we gained through this case study, as well as the model-predicted treatment schemes for individual patients. In the final section, we summarize the paper and discuss future work.

%% file: model.tex
\section{A Hybrid Model of Prostate Cancer Progression }\label{sec.model}

\begin{figure}[htb]
\centering
\includegraphics[scale=0.53]{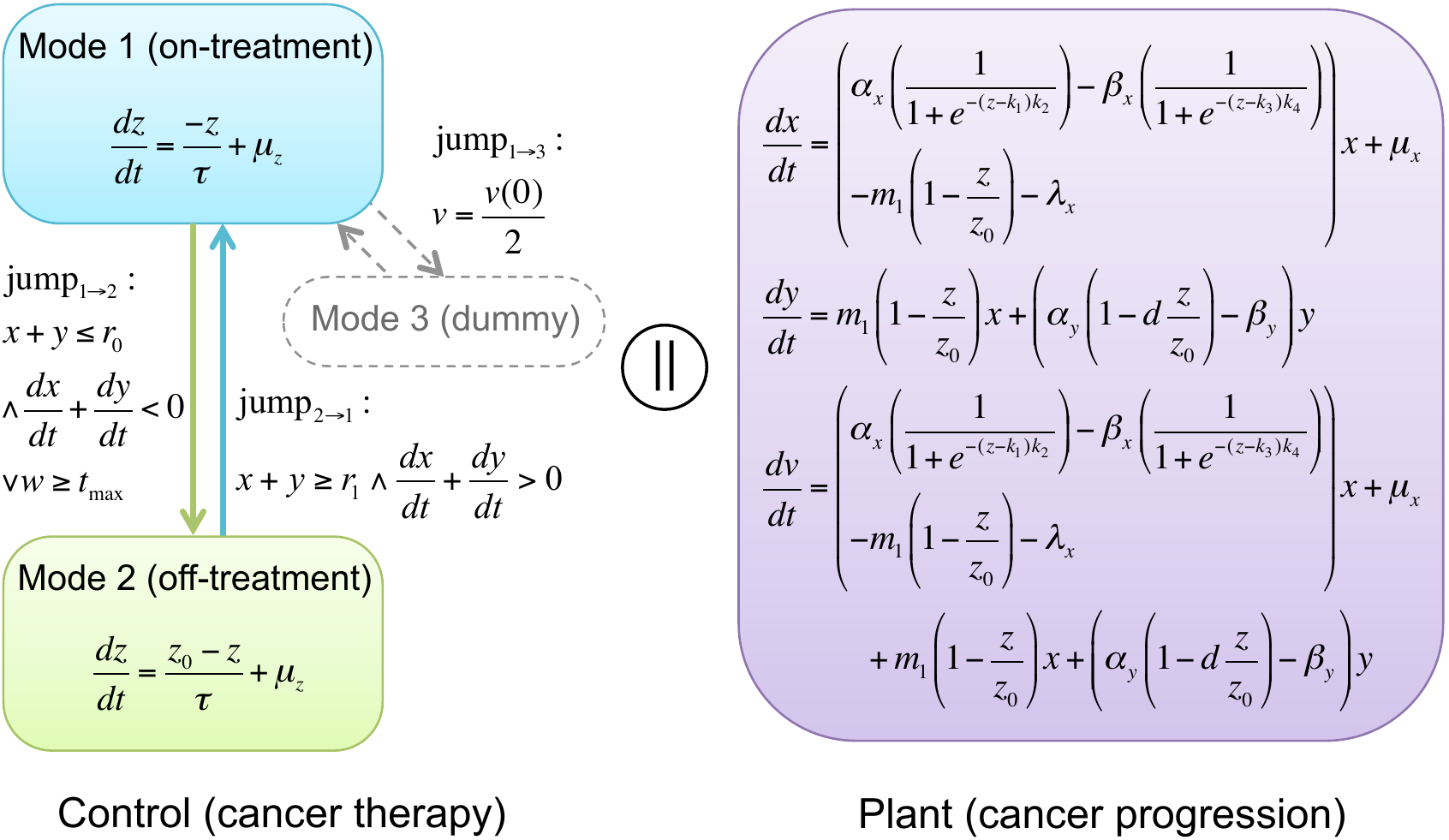}
\caption{A hybrid automaton model for prostate cancer hormone therapy. Symbol ``||'' denotes the parallel composition of the two automata.}
\label{pmodel}
 \vspace{-0.3cm}
\end{figure}

In this section, we propose a hybrid automata based model in order to reproduce the clinical observations \cite{bruchovsky06, bruchovsky07} of prostate cancer cell dynamics in response to the IAS therapy. It is known that the proliferation and survival of prostate cancer cells depend on the levels of androgens, specifically testosterone and 5$\alpha$-dihydrotestosterone (DHT).
Here we consider two distinct subpopulations of prostate cancer cells: hormone sensitive cells (HSCs) and castration resistant cells (CRCs). Androgen deprivation can lead to remarkable decreases of the proliferation and survival rates of HSCs, but also up-regulates the conversion from HSCs to CRCs, which will keep proliferating under low androgen level. %Figure \ref{progression} illustrates this cancer progression process and the corresponding hybrid automata model is shown in Figure \ref{pmodel}. 
The corresponding hybrid automata model is shown in Figure \ref{pmodel}.

%\begin{figure}[htb]
%\centering
%\includegraphics[scale=0.38]{fig-progression}
%\caption{Prostate cancer progression in response to (a) CAS and (b) IAS treatments.}
%\label{progression}
% %\vspace{-0.1cm}
%\end{figure}

% Reviewer#1
% Section 2 describes a 2 modes hybrid automaton, but later Section 4.2 presents a further modified automaton with 3 modes. It might be clearer to have the final version in Section 2 and Figure 2.

% Reviewer#2
% Figure 2: The dynamics for mode 1 and 2 appear to differ only in dz/dt and also differ depending on the jump direction. This figure could be greatly consolidated and the differences more clearly highlighted.

Our model is based on previous models developed by \cite{jackson04a,jackson04b,ideta08}. It takes into account the population of HSCs, the population of CRCs, as well as the serum androgen concentration, represented as $x(t)$, $y(t)$, and $z(t)$, respectively. In addition, it also includes the serum prostate-specific antigen (PSA) level $v(t)$, which is a commonly used biomarker for assessing the total population of prostate cancer cells. The model has two modes: \textit{on-treatment} mode and \textit{off-treatment} mode (note that the auxiliary Mode 3 will only be used in Section 4.2). Following \cite{ideta08}, in the off-treatment mode (Mode $2$), the androgen concentration is maintained at the normal level $z_0$ by homeostasis. In the on-treatment (Mode $1$), the androgen is cleared at a rate $\frac{1}{\tau}$. Further, we also introduce a basal androgen production rate $\mu_z$, in order to reproduce the measured basal testosterone levels in response to androgen suppression \cite{bruchovsky06, bruchovsky07}. 

\begin{figure}[htb]
\centering
\includegraphics[scale=0.45]{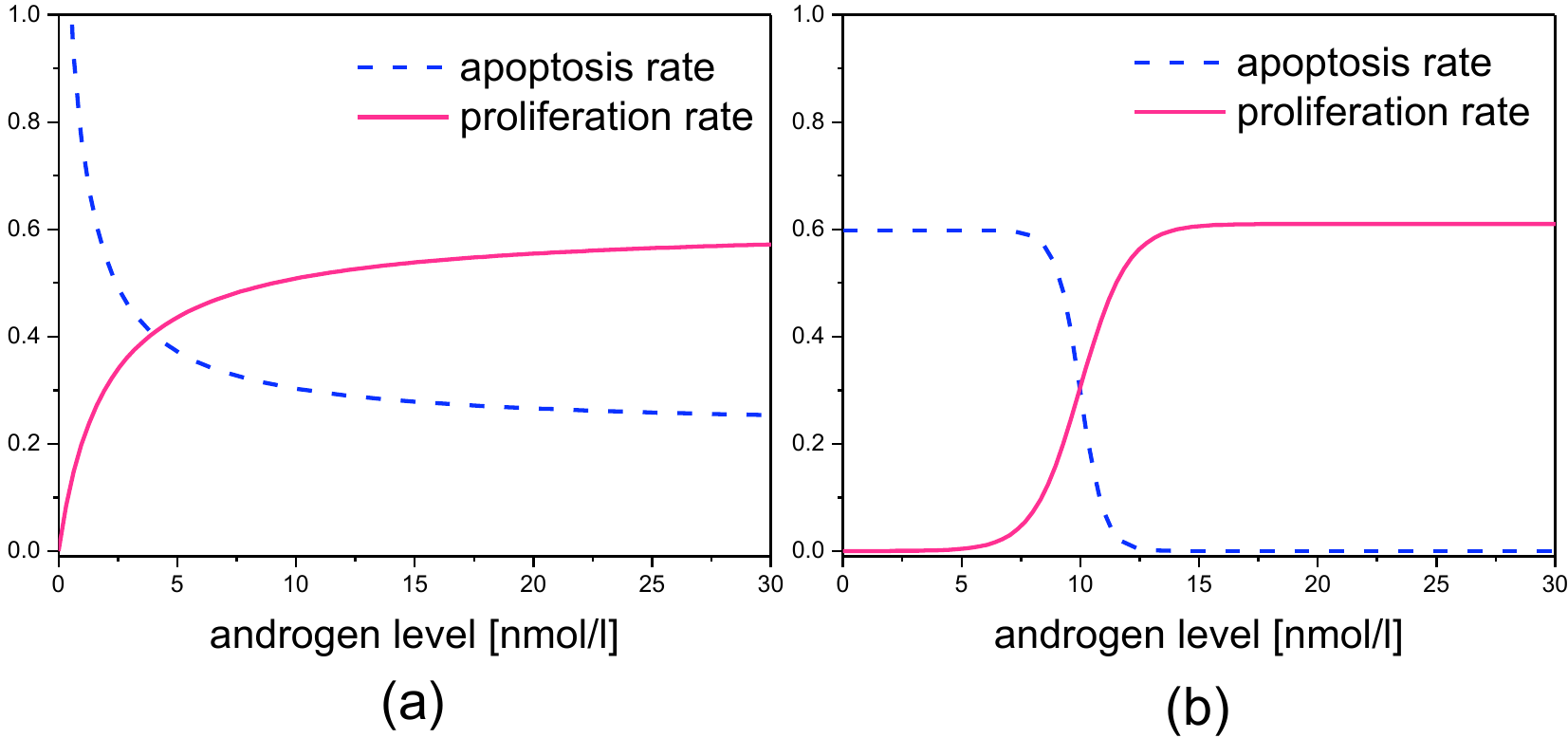}
\caption{Androgen response curves of (a) Ideta's model and (b) our model.}
\label{response}
% \vspace{-0.3cm}
\end{figure}

The net growth rate of $x(t)$ equals to $(prolif_{x}-apop_{x}-conv_{x})\cdot x(t)$, where $prolif_x$, $apop_x$ and $conv_x$ denote the proliferation, apoptosis and conversion rates, respectively. In previous studies such as \cite{jackson04a,jackson04b,ideta08}, the $prolif_x$ and $apop_x$ were modeled using Michaelis-Menten-like (MML) functions, in the form of $V_{max}+(1-V_{max})\frac{z(t)}{z(t)+K_{m}}$, where $V_{max}$ and $K_m$ are kinetic parameters. This approach will result in androgen response curves as shown in Figure \ref{response}(a). In particular, when one decreases the androgen level starting from the normal level, $prolif_x$ (or $apop_x$) begins to decrease (or increase) first slowly and then fast until a sufficiently low level of androgen is reached. However, this is inconsistent with the clinical observations presented in \cite{bruchovsky06, bruchovsky07}. The data show that for most of the patients, androgen suppression around normal level will induce an immediate decrease of the PSA level, which implies an fast decrease (or increase) of $prolif_x$ (or $apop_x$). Therefore, instead of the MML functions, we adopt sigmoid functions, in the form of  $\frac{1}{1+exp(-(z(t)-k_1)\cdot k_2)}$, to model $prolif_x$ and $apop_x$. The corresponding androgen response curves are shown in Figure \ref{response}(b). Following \cite{ideta08}, we model the conversion rate, proliferation rate and the apoptosis rate of $y(t)$ as $m_1(1-\frac{z(t)}{z_0})$, $\alpha_y(1-d\frac{z(t)}{z_0})$ and $\beta_y$, respectively. The PSA level $v$ (ng ml$^{-1}$) is defined as $v(t)=c_1\cdot x(t)+c_2\cdot y(t)$. 

The transitions between two modes depends on the values of $v$, ${dv}/{dt}$ and an auxiliary variable $w$, which measures the time taken in a mode. Specifically, for each patient we starts with mode $1$ to apply the treatment. When the PSA level drops to certain threshold $r_0$ or $w$ hits time out threshold $t_{max}$, the treatment will be suspended. When the PSA level is back to threshold $r_1$, the treatment will be resumed. Note that $w$ is associated with a dummy differential equation $\frac{dw}{dt}=1$ (not shown in Figure \ref{pmodel}). Its value will be reset to $0$ when the jump takes place.  

\begin{figure}[htb]
\centering
\includegraphics[scale=0.45]{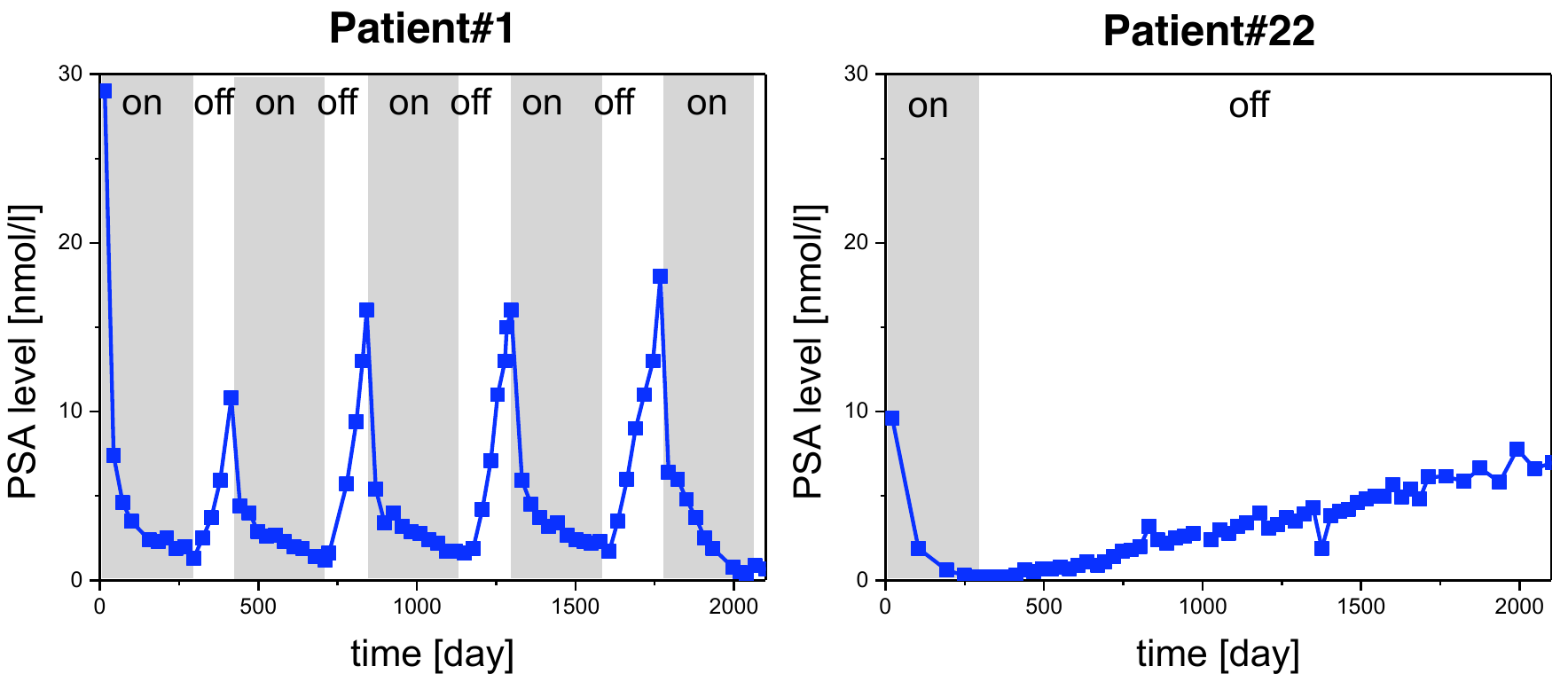}
\caption{The clinical data for PSA time serials.}
\label{data}
 \vspace{-0.3cm}
\end{figure}

We obtained the parameter values by fitting to patient PSA data reported in \cite{bruchovsky06, bruchovsky07}. Note that the patient-to-patient variability in terms of parameter values is significant. For example, Figure \ref{data} shows that the proliferation rate of Patient\#22 is much lower than the Patient\#1. The descriptions and a set of typical values (\ie~estimated from Patient\#1 data) of model parameters are listed in Table \ref{prostate}.

\begin{table}[ht]
\caption{Prostate cancer model parameter values\label{prostate}}
\centering
\small
\begin{tabular}{|c|c|c|}
\hline
Parameter  & Value & Remark  \\\hline
$\alpha_x$ & 0.0204 d$^{-1}$ & HSC proliferation \\
$\alpha_y$ & 0.0242 d$^{-1}$ & CRC proliferation  \\
$\beta_x$  & 0.0201 d$^{-1}$ & HSC apoptosis  \\
$\beta_y$  & 0.0168 d$^{-1}$ & CRC apoptosis \\
$k_1$     & 10.0 nM & HSC proliferation  \\
$k_2$     & 1.0 & HSC proliferation  \\
$k_3$     & 10.0 nM & HSC apoptosis  \\
$k_4$     &  2 & HSC apoptosis   \\
$m_1$     & 0.00005 d$^{-1}$ & HSC to CRC conversion  \\
$z_0$     & 12.0 nM & normal androgen level  \\
$\tau$     & 12.5 d & androgen degradation  \\
$\lambda_x$     & 0.01 d$^{-1}$ & HSC basal degradation \\
$\mu_x$     & 0.05 d$^{-1}$ & HSC basal production\\
$\mu_z$     & 0.02 d$^{-1}$ & Androgen basal production \\
\hline
\end{tabular}
% \vspace{-0.5cm}
\end{table}

%% file: method.tex
\section{Delta-Reachability Analysis}
%\section{Preliminaries}

%In this section, we present a computational framework for adaptively predicting the treatment schedules for prostate cancer patients. Figure \ref{} shows the flowchart of our approach. 

%In this section, we present the $\delta$-reachability analysis based techniques we have developed for hybrid

Hybrid automata are difficult to analyze. It has been shown that even simple reachability questions for hybrid 
systems with linear differential dynamics are undecidable \citep{henzinger96}. Therefore, in order to analyze our hybrid model of prostate cancer progression, we employed a $\delta$-reachability based framework \cite{liu14} which can sidesteps undecidability and allows the parameter synthesis problem to be relaxed in a sound manner and solved algorithmically. 

\subsection{Delta-Decisions}
The framework of $\delta$-complete decision procedures~\cite{gao12a} aims to solve first-order logic formula with arbitrary computable real functions, such as elementary functions and solutions of Lipschitz-continuous ODEs \citep{gao12b}. The answers returned by such procedures are either $\mathsf{unsat}$ or $\delta$-$\mathsf{sat}$. Here, $\mathsf{unsat}$ means the corresponding formula is verifiably false, while $\delta$-$\mathsf{sat}$ means a $\delta$-weakening version of the formula is true. In other words, $\delta$-decision procedures overcome undecidability issues by returning answers with one-sided $\delta$-bounded errors. Note that $\delta$ is an arbitrarily small positive rational chosen by the user. The algorithms for solving $\delta$-decision problems were described in our previous work \cite{gao12b,gao13} and were implemented in the dReal toolset \cite{dreal}. 

\subsection{Parameter identification}
Further, we have also proposed an encoding scheme which aimed to answer bounded reachability problems of hybrid automata with nontrivial invariants \cite{liu14}. This encoding enabled us to tackle the parameter identification problem by answering a $k$-step reachability question: ``Is there a parameter
combination for which the model reaches the goal region in $k$ steps?'' Essentially, we describe the set of states of interest (goal region) as a first-order logic formula and perform bounded model checking \cite{BMC} to determine reachability of these states. We then adapt an interval constraint propagation based algorithm to explore the parameter space and identify the sets of resulting parameters. If none exist, then the model is 
{\em unfeasible}. Otherwise, a witness (\ie, a value for each parameter) is returned. We have developed the dReach tool \cite{dreach} (\verb#http://dreal.cs.cmu.edu/dreach.html#) that automatically builds reachability formulas from a hybrid model and a goal description. Such formulas are then solved by the $\delta$-complete solver dReal \citep{dreal}.

For the interested readers, we refer to %Appendix (\url{http://www.cs.cmu.edu/~liubing/hscc15/}) and 
\cite{liu14} for more details on %$\delta$-decisions and 
$\delta$-reachability analysis based parameter identification.

%% file: results.tex
\section{Results}\label{sec.results}

We have implemented our prostate cancer progression model in the dReach's modeling language. The model files are available at \url{http://www.cs.cmu.edu/~liubing/hscc15/}. All the experiments reported below were done using a machine with two Intel Xeon E5-2650 2.00GHz processors and 32GB RAM. The precision $\delta$ was set to $10^{-3}$. 

\subsection{CRC proliferation dynamics}
Due to the lack of biomarkers distinguishing HSCs and CRCs \textit{in vivo}, the proliferation kinetics of CRCs in response to androgen is far from known. Three hypotheses, denoted as $H_1$, $H_2$ and $H_3$ have been proposed to describe the androgen-dependent CRC growth \cite{ideta08}, which are discriminated by the value of $d$ in the model, \ie:
% \vspace{-0.3cm}
\begin{itemize}
\item $H_1: d = 0$, the grow of CRCs is independent of $z(t)$;
\item $H_2: d = 1-\frac{\beta_y}{\alpha_y}$, CRCs cease growing when $z(t)=z_0$;
\item $H_3: d = 1$, CRCs decrease when $z(t)=z_0$.
\end{itemize} 
% \vspace{-0.3cm}

The Patient\#1 data presented in Figure 4 shows that with proper treatment schedules, it is possible to avoid his cancer relapse in years. We now show that only $H_3$ agrees with this observation. As the PSA level $v(t)$ reflects the total number of cancer cells and CRCs are responsible for recurrent cancer, we use two invariants: $0 \le v(t) \le 30$ and $0 \le y(t) \le 1$ to specify the property of ``no cancer relapse''. We then carried out $\delta$-reachability analysis to verify whether the invariants hold for each of the model candidates within a bounded time of $365$ days. Here the treatment schedule threshold parameters were provided as ranges: $r_0 \in [0, 7.99]$ (ng ml$^{-1}$) and $r_1 \in [8,15]$ (ng ml$^{-1}$).

%fixed as $r_0=4$ (ng ml$^{-1}$) and $r_1=10$ (ng ml$^{-1}$). The range of the initial concentration of androgen was given as $[15, 17]$ (nM).

\begin{figure}[t]
\centering
\includegraphics[scale=0.45]{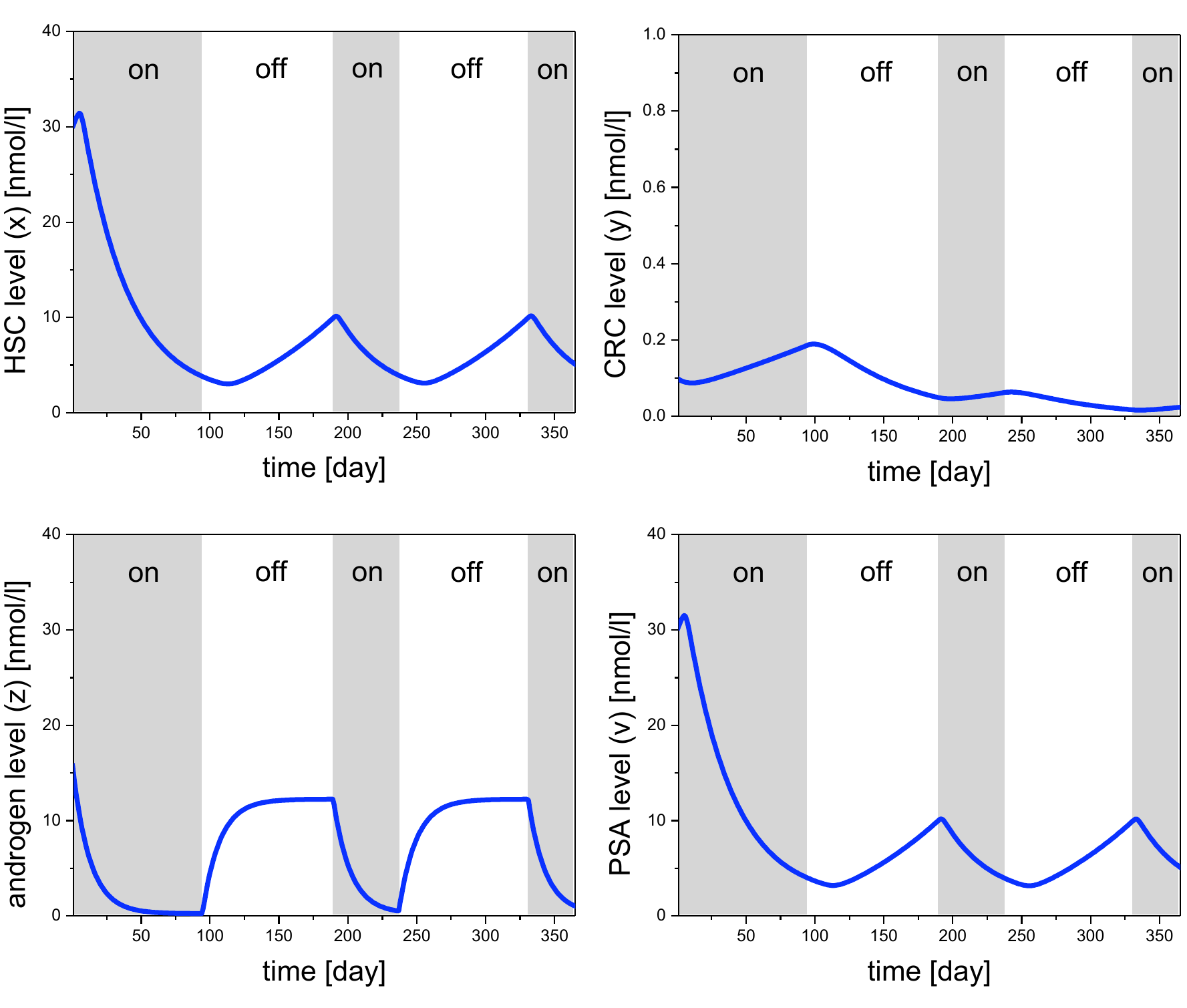}
\caption{The simulated witness trajectories of the $H_3$ model.}
\label{prostate-fig1}
%\vspace{-0.3cm}
\end{figure}

The $\mathsf{unsat}$ answers were returned for $H_1$ and $H_2$ (Run\#1 and Run\#2, Table \ref{runs}), indicating that they will always lead to cancer relapse no matter which 
%initial androgen concentration 
treatment schedule 
was chosen. In contrast, $\delta$-$\mathsf{sat}$ was returned for $H_3$ (Run\#3, Table \ref{runs}). Witness trajectories are shown in Figure \ref{prostate-fig1}), demonstrating that the cancer relapse can be avoided in a bounded time as observed experimentally \cite{ bruchovsky06,bruchovsky07}. The rest of the results in this paper were generated using $H_3$.

\subsection{Androgen-dependent HSC dynamics}

As mentioned in Section \ref{sec.model}, previous studies \cite{jackson04a,jackson04b,ideta08} modeled the androgen-dependent proliferation and apoptosis of HSCs using MML functions, while we use sigmoid functions. Here we show that the MML based approach is unable to reproduce an important dynamical property, but our model could. The patients' data in \cite{ bruchovsky06,bruchovsky07} show that the \textit{half-time} $t_{1/2}$ (\ie~the amount of time required for a quantity to fall to one half of its initial value) of PSA level under androgen suppression is often less than $60$ days. To specify this property, we introduced an auxiliary mode (Mode 3). If $v(t)=v(0)/2$, the system will jump from Mode 1 to Mode 3. Starting with Mode 1 and $20 \le x(0) \le 30$, we checked the reachability of a goal state with $0 \le w \le 60$ for both Ideta's model \cite{ideta08} and our model. The results show that  $\delta$-$\mathsf{sat}$ was returned for our model (Run\#4, Table \ref{runs}), while $\mathsf{unsat}$ was returned for Ideta's model (Run\#5, Table \ref{runs}), suggesting the superiority of sigmoid functions over MML functions in capturing HSC dynamics.

\begin{table}[!th]
  \centering
  \small
  \begin{tabular}{|c|c|c|c|c|}
    \hline
    \hline
    Run & Model & Initial State   & Result   & Time   \\
    \hline
    \hline
    1 & $H_1$ & $r_0 \in [0.0,7.99]$, $r_1 \in [8.0,15.0]$  & $\mathsf{unsat}$  &  3.94 \\
    2 & $H_2$ & $r_0 \in [0.0,7.99]$, $r_1 \in [8.0,15.0]$   & $\mathsf{unsat}$  &  5.26 \\
    3 & $H_3$ & $r_0 \in [0.0,7.99]$, $r_1 \in [8.0,15.0]$   & $\delta$-$\mathsf{sat}$ &  472 \\ 
    4 & $H_3$ & $x(0) \in [20.0,30.0]$   & $\delta$-$\mathsf{sat}$ &  10.1 \\
    5 & Ideta & $x(0) \in [20.0,30.0]$   & $\mathsf{unsat}$ &  0.5 \\           
    6 & $H_3$ & $r_0 \in [0.0,7.99]$, $r_1 \in [8.0,15.0]$   & $\delta$-$\mathsf{sat}$ &  526 \\ 
    7 & $H_3$ & $r_0 \in [0.0,7.99]$, $r_1 \in [8.0,15.0]$   & $\mathsf{unsat}$ &  0.3 \\ 
    8 & $H_3$ & $r_0 \in [0.0,7.99]$, $r_1 \in [8.0,15.0]$   & $\delta$-$\mathsf{sat}$ &  28 \\ 
    9 & $H_3$ & $r_0 \in [0.0,7.99]$, $r_1 \in [8.0,15.0]$   & $\delta$-$\mathsf{sat}$ & 203 \\ 
    \hline
    \hline
  \end{tabular}
  \caption{\small
  Experimental results.
    Result - bounded model checking result,
    Time - CPU time (s),
    $\delta=10^{-3}$,
    Model parameters used in Run\#1-5 are listed in Table 1,  model parameters used in Run\#6-9 are listed in Table 3.
    %Trace = Size of the ODE trajectory
}\label{runs}
% \vspace{-0.3cm}
\end{table}
%}

\begin{table*}[t]
\caption{Estimated personalized parameters and suggested treatment schemes \label{prostate2}}
\centering
\small
\begin{tabular}{|c|c|c|c|c|}
\hline\hline
Parameter  & Patient\#1 & Patient\#11 & Patient \# 15 & Patient\#26  \\\hline\hline
$\alpha_x$ & 0.0204 d$^{-1}$ & 0.0204 d$^{-1}$ & 0.0213 d$^{-1}$& 0.0197 d$^{-1}$ \\
$\alpha_y$ & 0.0242 d$^{-1}$ &  0.0242 d$^{-1}$&  0.0242 d$^{-1}$&  0.0242 d$^{-1}$  \\
$\beta_x$  & 0.0201 d$^{-1}$ & 0.02 d$^{-1}$ & 0.01 d$^{-1}$& 0.0175 d$^{-1}$ \\
$\beta_y$  & 0.0168 d$^{-1}$ &0.0158 d$^{-1}$ &0.0168 d$^{-1}$ &0.0168 d$^{-1}$  \\
$k_1$     & 10.0 nM & 7.0 nM & 7.0 nM& 10.0 nM  \\
$k_2$     & 1.0 & 1.0 & 1.0 & 1.0 \\
$k_3$     & 10.0 nM & 7.0 nM & 7.4 nM& 10.0 nM  \\
$k_4$     &  2 & 2 & 2 & 2   \\
$m_1$     & 0.00005 d$^{-1}$ &  0.00005 d$^{-1}$ &  0.00005 d$^{-1}$ &  0.00005 d$^{-1}$  \\
$z_0$     & 12.0 nM & 9.0 nM & 8.0 nM & 12.0 nM  \\
$\tau$     & 12.5 d & 12.5 d & 12.5 d & 12.5 d  \\
$\lambda_x$     & 0.01 d$^{-1}$ & 0.0121 d$^{-1}$ & 0.01 d$^{-1}$ & 0.01 d$^{-1}$ \\
$\mu_x$     & 0.05 d$^{-1}$ & 0.06 d$^{-1}$ & 0.02 d$^{-1}$ & 0.03 d$^{-1}$\\
$\mu_z$     & 0.02 d$^{-1}$ & 0.02 d$^{-1}$  & 0.02 d$^{-1}$ & 0.02 d$^{-1}$  \\
\hline\hline
Scheme     & $r_0=5.2$, $r_1=10.8$ & N.A & $r_0=1.9$, $r_1=8.0$ & $r_0=4.6$, $r_1=10.7$ \\
\hline\hline
\end{tabular}
 \vspace{-0.3cm}
\end{table*}

\subsection{Personalized therapy design}
\begin{figure}[htb]
\centering
\includegraphics[scale=0.43]{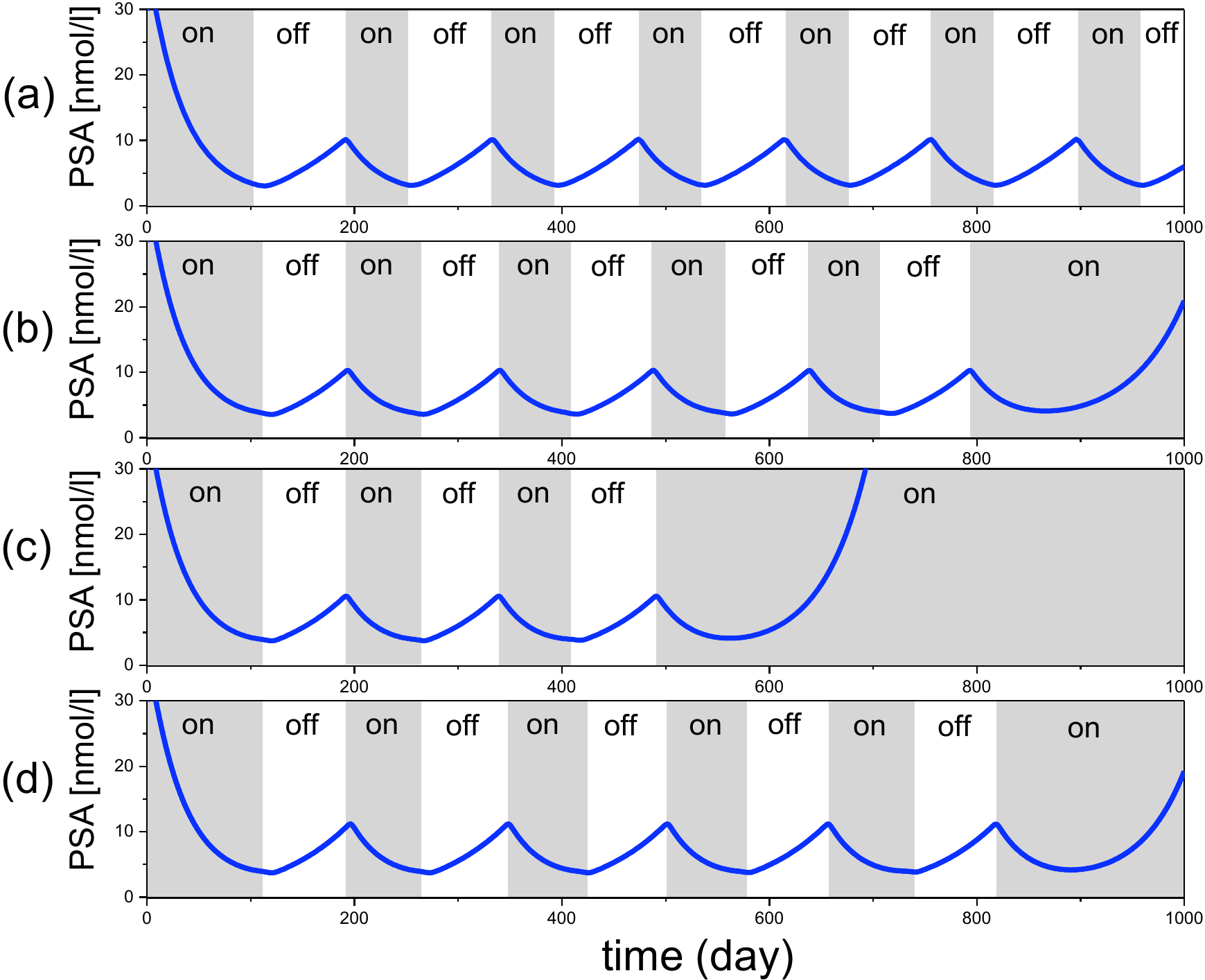}
\caption{Simulated PSA profiles of three mock patients with different parameters. (a) Patient A: $\alpha_y=0.0242$, $\beta_y=0.0168$, $m_1=0.00005$, $z(0)=12$, $r_0=4$, $r_1=10$ (b) Patient B: $\alpha_y=0.0328$, $\beta_y=0.013$, $z(0)=13$, $m_1=0.0001$, $r_0=4$, $r_1=10$ (c) Patient C: $\alpha_y=0.0426$, $\beta_y=0.189$, $m_1=0.00005$, $z(0)=15$, $r_0=4$, $r_1=10$ (d) Patient C with $r_0=4$, $r_1=10.6$.}
\label{patients}
% \vspace{-0.5cm}
\end{figure}

We next apply $\delta$-reachability analysis to design treatment schemes for individual patients. The parameter values shown in Table \ref{prostate} were estimated by fitting the data of Patient\#1. Since the IAS response of Patient\#1 is typical, we treated its parameter values as the baseline values. As we demonstrated in Figure \ref{data}, the values of some parameters vary among patients. Such variability may significantly affect the hormone therapy responses. 
For example, Figure \ref{patients}(a-c) illustrates the PSA dynamics of $3$ mock patients with different personalized parameters under the same IAS treatment scheme ($r_0=4$, $r_1=10$). IAS prevents the relapse for Patient A and delays the relapse for Patient B, but does not help Patient C. Figure \ref{patients}(d) shows that, by modifying the IAS scheduling parameters $r_0$ and $r_1$, the relapse of Patient C can be avoided or delayed. 

Given the parameter values of an particular patient, we can design a treatment scheme, which might help him avoid cancer relapse with bounded time by solving the following parameter identification problem: (i) set the ranges of scheduling parameters as $r_0 \in [0,7.99]$ (nM) and $r_1 \in [8,15]$; (ii) check if $H_3$ can reach the goal state without violating the ``no cancer relapse'' invariants within $1$ year. If $\mathsf{unsat}$ was returned, it means that androgen suppression therapy is not suitable for the patient. The patient then has to resort to other kinds of therapeutic interventions. Otherwise, when the $\delta$-$\mathsf{sat}$ answer is returned, a treatment scheme containing feasible values of $r_0$ and $r_1$ will also be returned, which could help in preventing or delaying the relapse within bounded time. Note that if $r_0=0$ is returned, it implies that the CAS scheme, instead of IAS scheme, might be more suitable for the patient.

The personalized parameters of individual patients can be obtained by collectively fitting the available experimental data. We tested our method on real patients data collected by \cite{bruchovsky07}\footnote{Data available at \url{http://www.nicholasbruchovsky.com/clinicalResearch.html}.}. The parameter values for each randomly selected patient were estimated by fitting the model to the PSA time serials data under the IAS therapy using an evolutionary strategy search, which is capable of estimating parameters from noisy biological data \cite{moles03}.

\begin{figure}[htb]
\centering
\includegraphics[scale=0.45]{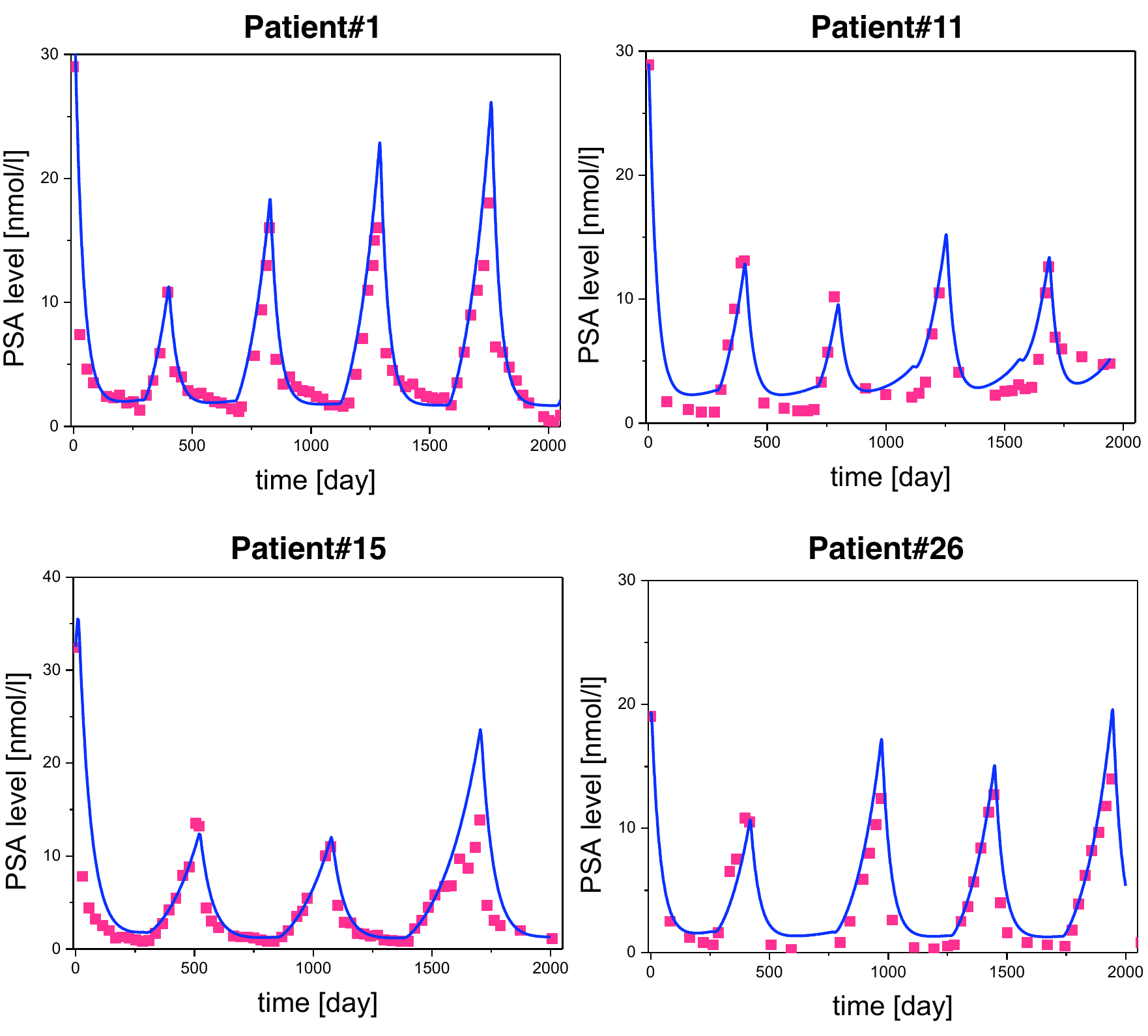}
\caption{Model prediction vs. experimental data.}
\label{fitting}
\vspace{-0.3cm}
\end{figure}

As an example, Figure \ref{fitting} shows the comparison between model predictions and the experimental data of PSA and androgen levels for Patient\#1, Patient\#11, Patient\#15, and Patient\#26. We then predicted the treatment schemes for the future year using $\delta$-reachability analysis (Run\#6 for Patient\#1, Run\#7 for Patient\#15, Run\#8 for Patient\#26 and Run\#9 for Patient\#11, Table \ref{runs}). The results are summarized in Table \ref{prostate2}. Note that for Patient\#11, $\mathsf{unsat}$ was returned, implying that no suitable treatment schemes were identified. This might be due to the raised population size of CRCs in the late phase of clinical trails. 

% reviewer#4
% some discussion on the experience of using dReal in this case study would be interesting. For example, is the choice of delta critical ? Why are we observing such variability in the computational times of table 2 ? etc.

% [Bing]: address these in the journal version?

%% file: conclusion.tex
\section{Conclusion}
\label{sec:Conclusion}

We have proposed a hybrid model to study the prostate cancer cell dynamics in response to hormone therapy. Using $\delta$-reachability analysis, we obtained interesting biological insights into the prostate cancer heterogeneity. We also developed a $\delta$-decisions based computational framework for predicting patient-specific treatment schedules. We have demonstrated the applicability of our method with the help of real clinical datasets. Our study explored the possibilities of using formal methods to tackle quantitative systems pharmacology problems. Our results also highlighted $\delta$-reachability analysis as a potent technique in this line of research. 

Experimental validation of our method might require years of clinical studies, which is beyond the scope of this case study. It is worth noting that our therapy design framework is generic and can be applied to other settings, for example, predicting the radiation dosing schedules for brain cancer \cite{leder14}. Furthermore, another interesting direction is to extend our model and framework to capture cancer hallmarks and/or to take into account the stochasticity of a cellular environment. In this respect, the cancer hybrid automata formalization \cite{bud14} and the probabilistic/statistical analysis techniques in \cite{liu11,liu13,liu12bioinfo} might offer helpful pointers.